\documentclass[12pt]{iopart}

\usepackage{graphicx}
\usepackage{color}
\usepackage{iopams}
\usepackage{multirow}

\begin{document}

\title{Fitting N$^{3}$LO pseudopotentials through central plus tensor Landau parameters}

\author{D. Davesne}
\address{Universit{\'e} de Lyon, F-69622 Lyon, France, \\
Universit\'e Lyon 1, Villeurbanne;  CNRS/IN2P3, UMR5822, Institut de Physique Nucl\'eaire de Lyon}

\author{A. Pastore}
\address{Institut d'Astronomie et d'Astrophysique, CP 226, Universit\'e Libre de Bruxelles, B-1050 Bruxelles, Belgium}

\author{J. Navarro}
\address{IFIC (CSIC-Universidad de Valencia), Apartado Postal 22085, E-46.071-Valencia, Spain}

\date{\today}

\begin{abstract}
Landau parameters determined from phenomenological finite-range  interactions are used to get an estimation of N$^3$LO pseudopotentials parameters. The parameter sets obtained in this way are shown to lead to consistent results concerning saturation properties. The uniqueness of this procedure is discussed, and an estimate of the error induced by the truncation at N$^3$LO is given.
\end{abstract}

\maketitle

\section{Introduction}

The search for a general nuclear energy density functional (EDF) is currently attracting intense efforts \cite{ben03,kor10,kor13,sad13,gor13,was12}. 
In this paper we focus on the method initiated in \cite{car08}, where the nuclear EDF was expanded 
in higher-order derivatives of densities up to the next-to-next-to-next-to-leading order (N$^3$LO). 
Those results were completed in \cite{rai11} by deriving the expression of the effective pseudopotential in relative 
momenta up to the same order. In that language, the standard Skyrme pseudopotential corresponds to the N$^1$LO, plus density dependent terms. 
In principle, the introduction of higher orders could improve the description of nuclear properties once the related
parameters are fixed. At present the only application of the N$^{3}$LO has been presented in \cite{car10}, 
within the context of the density matrix expansion (DME). The parameters obtained in this way can be considered 
as a starting point for a complete minimization procedure \cite{kor10,sto10}. 
 
In a recent work \cite{dav13}, we have given the explicit form in Cartesian basis of the N$^2$LO effective 
pseudopotential compatible with all required symmetries, and especially with gauge invariance. We have also 
suggested that numerical values of the Landau parameters could be used to put explicit constraints on the values of the coupling constants. The main objective of the present paper is to explore the possibility of using Landau parameters
to determine the pseudopotential parameters.  As we shall show, although the solution is not unique it is possible 
to reasonably fix their values, which could provide an alternative starting point to the DME one.

The plan of the paper is as follows. 
In Section \ref{sect-param} we extend the results of \cite{dav13} by presenting the N$^3$LO pseudopotential in Cartesian basis. In Section \ref{sect-landau} we relate the N$^3$LO pseudopotential parameters to the Landau parameters, and discuss problems and cures when fixing the former from the latter. In Section \ref{sect-results} we determine two N$^3$LO pseudopotentials from the Landau parameters calculated from two effective finite-range two-body interactions. The conclusions of the present study are formulated in Section \ref{sect-conclusions}. Some interesting formulae are given in an Appendix.  

\section{The Skyrme pseudopotential up to N$^3$LO} \label{sect-param}

The mathematical construction of a pseudopotential is equivalent to the construction of
scalars with relative momenta and Pauli matrices as basic ingredients. The general expression up to N$^3$LO can be written as:
%%%
\begin{eqnarray}
\hat{V}_{\rm Sk} = \hat{V}^{(0)}_{\rm Sk} + \hat{V}^{(2)}_{\rm Sk} + \hat{V}^{(4)}_{\rm Sk} + \hat{V}^{(6)}_{\rm Sk} \, .
\end{eqnarray}
%%%
The zeroth- plus second- order represents the  Skyrme pseudopotential involving the parameters 
$t_0, x_0, t_1, x_1, t_2, x_2$ as well as the tensor ones $t_e, t_o$:
%%%
\begin{eqnarray}
\fl \hat{V}^{(0)}_{\rm Sk} &=& t_0 (1+x_0 P_{\sigma}) \label{V-zero}\, , \\
\fl \hat{V}^{(2)}_{\rm Sk} &=& 
\frac{1}{2} t_1 (1+x_1 P_{\sigma}) ({\bf k}^2 + {\bf k'}^2) + t_2 (1+x_2 P_{\sigma}) ({\bf k} \cdot {\bf k'})  
+ \frac{1}{2} t_e T_e({\bf k'},{\bf k})  + \frac{1}{2} t_o T_o({\bf k'},{\bf k}) \label{V-two} \, , 
\end{eqnarray}
%%%
with the usual definitions ${\bf k}=(\overrightarrow{\nabla}_1-\overrightarrow{\nabla}_2)/2 i$ and
${\bf k'}=-(\overleftarrow{\nabla}_1-\overleftarrow{\nabla}_2)/2 i$. The above  Skyrme parameters are related to the N$^1$LO coupling constants EDF as described in \cite{rai11}. 
The operators $T_e$ and $T_o$, respectively even and odd under parity transformation, are defined as:
%%%
\begin{eqnarray}
\fl T_e({\bf k'},{\bf k}) &=& 3 (\vec \sigma_1 \cdot {\bf k'}) (\vec \sigma_2 \cdot {\bf k'}) 
+ 3 (\vec \sigma_1 \cdot {\bf k}) (\vec \sigma_2 \cdot {\bf k}) 
- ({\bf k'}^2 + {\bf k}^2) (\vec \sigma_1 \cdot \vec \sigma_2), \\
\fl T_o({\bf k'},{\bf k}) &=& 3 (\vec \sigma_1 \cdot {\bf k'}) (\vec \sigma_2 \cdot {\bf k}) 
+ 3 (\vec \sigma_1 \cdot {\bf k}) (\vec \sigma_2 \cdot {\bf k'}) 
- 2 ({\bf k'} \cdot {\bf k}) (\vec \sigma_1 \cdot \vec \sigma_2).
\end{eqnarray} 
%%%
In these expressions, as in the coming analogous ones, a $\delta({\bf r}_1-{\bf r}_2)$ function is to be understood, 
but has been omitted for the sake of clarity. 

The fourth order pseudopotential was deduced in \cite{dav13}, and written by keeping a close analogy with the second order structure:
%%%
\begin{eqnarray}
\fl \hat V ^{(4)}_{\rm Sk} &=& 
 \frac{1}{4} t_1^{(4)} (1+x_1^{(4)} P_{\sigma}) \left[({\bf k}^2 + {\bf k'}^2)^2 + 4 ({\bf k'} \cdot {\bf k})^2\right] \nonumber \\
\fl &+& t_2^{(4)} (1+x_2^{(4)} P_{\sigma}) ({\bf k'} \cdot {\bf k}) ({\bf k}^2 + {\bf k'}^2) \nonumber \\
\fl &+&  t_e^{(4)} \left[  ({\bf k}^2+{\bf k'}^2) T_e({\bf k'},{\bf k})  
+  2 ({\bf k'} \cdot {\bf k}) T_o({\bf k'},{\bf k})  \right] \nonumber \\
\fl &+& t_o^{(4)} \left[ ({\bf k}^2+{\bf k'}^2) T_o({\bf k'},{\bf k})  
+  2 ({\bf k'} \cdot {\bf k}) T_e({\bf k'},{\bf k}) \right] , 
\label{V-four}
\end{eqnarray}
%%%
The relations between the parameters and the N$^2$LO coupling constants of \cite{rai11} are given in \cite{dav13}.  We correct here a misprint in \cite{dav13} regarding the parameter $t_o^{(4)}$, whose right expression is  
$t_o^{(4)} = -\frac{ C_{11,22}^{33}}{12\sqrt{7}}$. We also correct the expression of the term proportional to $t_o^{(4)}$ which is now symmetric with respect to the term proportional to $t_e^{(4)}$.

We have extended these results to the next order. Starting from the general N$^3$LO  pseudopotential derived in \cite{rai11} we have obtained its explicit form in the more familiar Cartesian basis, constraining it to be 
gauge invariant. 
Although this symmetry is not explicitly required from basic principles, there is some current discussion about the 
necessity of imposing it in general, since it has been shown \cite{rai11b} that gauge invariance is equivalent to 
continuity equation for local potentials. The continuity equation is of particular interest in view of using such a 
pseudopotential for calculations of the time evolution of a quantal system. For this reason, and from the fact 
that a local potential is automatically gauge invariant \cite{doba95,bla86}, we only consider pseudopotentials
 which are gauge invariant at all orders. 
Following the method explained in \cite{dav13}, we have worked out the 6th order, which we write as:
%%%
\begin{eqnarray}
\fl    \hat V ^{(6)}_{\rm Sk} &=& 
\frac{t_1^{(6)}}{2}(1+x_1^{(6)}P_\sigma) ({\bf k}^2 + {\bf k'}^2)\left[({\bf k}^2 + {\bf k'}^2)^2 + 12 ({\bf k'} \cdot {\bf k})^2\right] \nonumber \\
\fl  &+& t_2^{(6)}(1+x_2^{(6)}P_\sigma)
({\bf k'} \cdot {\bf k}) \left[3({\bf k} ^2 + {\bf k'}^2)^2 +4({\bf k'} \cdot {\bf k})^2\right] \nonumber \\ 
\fl &+& t_e^{(6)} \left[ \left(\frac{1}{4}({\bf k} ^2 + {\bf k'}^2)^2 +({\bf k'} \cdot {\bf k})^2\right) T_e({\bf k'},{\bf k})
+ ({\bf k} ^2 + {\bf k'}^2) ({\bf k'} \cdot {\bf k})  T_o({\bf k'},{\bf k}) \right] \nonumber \\
\fl &+& t_o^{(6)} \left[\left(\frac{1}{4}({\bf k} ^2 + {\bf k'}^2)^2 +({\bf k'} \cdot {\bf k})^2\right) T_o({\bf k'},{\bf k})
+ ({\bf k} ^2 + {\bf k'}^2) ({\bf k'} \cdot {\bf k})  T_e({\bf k'},{\bf k}) \right] \, .
\label{V-six}
\end{eqnarray}
%%%
In the same way as we did for the 4th order, the definition of 6th order parameters has been chosen 
such that their contributions to the equation of state and Landau parameters maintain a close analogy with 
those of 2nd order, as shown below. This is the first noticeable result of this paper. The relations of these 6th 
order parameters to the N$^3$LO coupling constants of \cite{rai11} are the following: 
%%%
\begin{eqnarray*}
\fl \frac{1}{2} t_1^{(6)} = \frac{\sqrt{3} C_{22,00}^{42} + C_{22,20}^{42}}{72\sqrt{5}} &,&
\frac{1}{2} t_1^{(6)}  x_1^{(6)} =  -\frac{C_{22,20}^{42}}{36\sqrt{5}} \, , \\ 
\fl t_2^{(6)} = \frac{3 C_{33,00}^{33} + \sqrt{3} C_{33,20}^{33}}{12\sqrt{7}} &,&
t_2^{(6)}  x_2^{(6)} =  -\frac{C_{33,20}^{33}}{6}\sqrt{\frac{3}{7}} \, , \\
\fl t_e^{(6)} = - \frac{C_{22,22}^{44} }{18}  &,&
t_o^{(6)} = -\frac{ C_{11,22}^{53}}{18}\sqrt{\frac{3}{7}} \, . 
\end{eqnarray*}
  
As discussed in \cite{dav13}, the spin-orbit contribution appears only at the N$^1$LO level. We have not written it in (\ref{V-two}) because it does not contribute to the Landau parameters. Notice that no density dependent terms appear in this procedure, as they are originated from three-, four- ... body interactions~\cite{vau72}. In practice they are added by hand as phenomenological terms, being in general the same as the standard Skyrme interaction \cite{cha98,cha98b}. 

\section{N$^3$LO parameters and Landau parameters} \label{sect-landau}

In the Landau-Migdal approximation  it is assumed that the low-energy excitations of the system are described 
by putting the interacting particles and holes on the Fermi surface, so that the only variable that remains is the relative angle $\theta$ between the initial and final momenta. The p-h interaction is thus a contact interaction, which 
is expanded in Legendre polynomials with argument $\cos \theta = ( {\mathbf {\hat k}_1} \cdot {\mathbf {\hat k}_2}  )$. It includes spin and isospin degrees of freedom, 
and the general form for symmetric nuclear matter (SNM) adopted here reads:
%%%
\begin{eqnarray}
\label{landau-Vph}
 V_{ph}  &=&  \sum_{\ell} \,
\bigg\{ f_{\ell} + f_{\ell}' \, (\btau_1 \cdot \btau_2 ) 
+ \left[ g_{\ell}+ g_{\ell}' \, (\btau_1 \cdot \btau_2) 
\right] (\bsigma_1 \cdot \bsigma_2) \nonumber \\
&& \quad  + \left[ h_{\ell} + h_{\ell}' \, (\btau_1 \cdot \btau_2 ) \right]  \frac{{\mathbf k}_{12}^{2}}{k_{F}^{2}} \, S_{12}
\bigg\} \, P_{\ell} ( {\mathbf {\hat k}_1} \cdot {\mathbf {\hat k}_2}  ) \, ,
\end{eqnarray}
%%%
where $f_{\ell}, f'_{\ell}, \dots$ are the Landau parameters, ${\mathbf k}_{12} = {\mathbf k}_1 - {\mathbf k}_2$, and 
$S_{12}= 3( \hat{\mathbf{k}}_{12} \cdot \bsigma_1)( \hat{\mathbf{k}}_{12} \cdot \bsigma_2 )- ( \bsigma_1 \cdot \bsigma_2) $ is the usual tensor operator. Excitations are characterized by the spin and isospin quantum numbers $(S, I)$. In the case of pure neutron matter (PNM) the p-h interaction only depends on the spin quantum number. Eq. (\ref{landau-Vph}) is then modified by dropping the coefficients $f'_{\ell}, g'_{\ell}, h'_{\ell}$ and using the notation $f^{(n)}_{\ell}, g^{(n)}_{\ell}, h^{(n)}_{\ell}$ for the remaining ones. In the following, we shall use a single symbol $(\alpha)$ to indicate the relevant spin-isospin quantum numbers. With this notation, the Landau parameters $f_{\ell}, f'_{\ell}, g_{\ell}, g'_{\ell}$ will be written as $f_{\ell}^{(\alpha)}$, with $(\alpha) = (0,0), (0,1), (1,0), (1,1)$, respectively. 

Our objective is to write the pseudo-potential parameters as linear combinations of Landau parameters. 
Let us consider separately central and tensor parameters.
 
\subsection{Central parameters}
Up to N$^3$LO, the central Landau parameters can be written as:
%%%
\begin{equation} \left.
\begin{array}{rl}
 f_0^{(\alpha)} =&  \frac{1}{4} L_0[f^{(\alpha)}] + \frac{1}{8} k_F^2 L_2[f^{(\alpha)}] + \frac{1}{6} k_F^4 L_4[f^{(\alpha)}] + k_F^6 L_6[f^{(\alpha)}]   \\ 
 f_1^{(\alpha)} =& - \frac{1}{8} k_F^2 L_2[f^{(\alpha)}] - \frac{1}{4} k_F^4 L_4[f^{(\alpha)}] - \frac{9}{5} k_F^6 L_6[f^{(\alpha)}]  \\ 
 f_2^{(\alpha)} =&   \frac{1}{12} k_F^4 L_4[f^{(\alpha)}] + k_F^6 L_6[f^{(\alpha)}]  \\
f_3^{(\alpha)} =&  - \frac{1}{5} k_F^6 L_6[f^{(\alpha)}]  
\end{array} \right\} 
\label{f0-3} 
\end{equation}
%%%
where the $L_n$ terms are combinations of the pseudopotential parameters. One has to keep in mind that no 
density-dependent terms are considered for the moment. The explicit combinations are:
%%%
\begin{equation} \left.
\begin{array}{rl}
L_0[f] =& 3 t_0  \\
L_0[g] =& -  t_0 (1-2 x_0)  \\
L_0[f'] =&  -  t_0 (1+2 x_0) \\
L_0[g'] =& -  t_0 
 \end{array} \right\}
 \label{Leq0}
 \end{equation}
and for $n \ge 2$:
\begin{equation} \left.
\begin{array}{rl}
L_n[f] =& 3 t_1^{(n)} + (5+4x_2^{(n)}) t_2^{(n)}  \\ 
L_n[g] =&  - (1-2x_1^{(n)}) t_1^{(n)} + (1+2x_2^{(n)}) t_2^{(n)}  \\ 
L_n[f'] =& -(1+2x_1^{(n)}) t_1^{(n)} + (1+2x_2^{(n)}) t_2^{(n)}  \\ 
L_n[g'] =& - t_1^{(n)} +  t_2^{(n)} 
 \end{array} \right\}
 \label{Lgt0}
 \end{equation}

From equations (\ref{f0-3}) one can first write $L_n$ as combinations of Landau parameters. Then, including the result in the systems (\ref{Leq0}) and (\ref{Lgt0}), one gets the pseudopotential parameters in terms of those combinations.  
The system (\ref{Lgt0}) has a unique solution, and the Skyrme parameters for $n \ge 2$ can be written as a linear combination of the symmetric nuclear matter Landau parameters with $\ell \ge 1$ as~:
%%%
\begin{eqnarray*}
\fl - \frac{8}{5} k_F^6 t_1^{(6)} &=& f_3 - g_3 - f'_3 - 3 g'_3\, , \\
\fl - \frac{4}{5} k_F^6 t_1^{(6)} x_1^{(6)} &=&  g_3 - f'_3  \, ,\\
\fl - \frac{8}{5} k_F^6 t_2^{(6)} &=& f_3 - g_3 - f'_3 + 5 g'_3\, , \\
\fl - \frac{4}{5} k_F^6 t_2^{(6)} x_2^{(6)} &=&  g_3 + f'_3 - 2 g'_3 \, ,
\end{eqnarray*}
\begin{eqnarray*}
\fl \frac{2}{3} k_F^4 t_1^{(4)} &=& f_2-g_2-f'_2-3g'_2 + 5 \left( f_3 - g_3 - f'_3 - 3 g'_3 \right)\, , \\
\fl \frac{1}{3} k_F^4 t_1^{(4)} x_1^{(4)} &=&  g_2-f'_2 + 5 \left( g_3 - f'_3 \right)\, , \\
\fl \frac{2}{3} k_F^4 t_2^{(4)} &=& f_2 - g_2 - f'_2 + 5 g'_2 + 5 \left( f_3 -  g_3 -  f'_3 + 5 g'_3 \right) \, ,\\
\fl \frac{1}{3} k_F^4 t_2^{(4)} x_2^{(4)} &=&  g_2 + f'_2 - 2 g'_2 +5  \left(g_3 +  f'_3 - 2 g'_3 \right)\, ,
\end{eqnarray*}
\begin{eqnarray*}
\fl - k_F^2 t_1 &=& f_1-g_1-f'_1-3g'_1+ 3 \left( f_2-g_2-f'_2-3g'_2\right) + 6 \left(f_3 - g_3 - f'_3 - 3 g'_3 \right)\, ,\\
\fl - \frac{1}{2} k_F^2 t_1 x_1 &=& g_1-f'_1 + 3 \left(g_2-f'_2\right) + 6 \left( g_3 -f'_3 \right)\, , \\
\fl - k_F^2 t_2 &=&  f_1 - g_1 - f'_1 + 5 g'_1+3 \left( f_2 - g_2 - f'_2 + 5 g'_2 \right)+ 6 \left( f_3 - g_3 - f'_3 + 5 g'_3 \right)\, , \\
\fl - \frac{1}{2} k_F^2 t_2 x_2 &=& g_1 + f'_1 - 2 g'_1+ 3 \left(g_2 + f'_2 - 2g'_2 \right) +6 \left( g_3 + f'_3 - 2 g'_3 \right) \,.
\end{eqnarray*}

Things are more complicated for $n=0$, because the four equations (\ref{Leq0}) are not linearly independent. 
It has no solution  unless some relation between Landau parameters are fulfilled. 
For instance, from the last three equations of system  (\ref{Leq0}) one  finds $L_0[g]+L_0[f']=2L_0[g']$,  
 which in turn implies:
 %%%
\begin{eqnarray}
\delta_3 =  \sum_{\ell=0}^{3}\left( g_{\ell} + f'_{\ell} - 2  g'_{\ell} \right) = 0 \, .
\label{eq-SNM}
\end{eqnarray}
%%%
One would be surprised that this relation holds in general. Indeed, in \ref{Lan-Cen} we show that 
in the case of a finite-range two-body central interaction, the quantity $\delta_{\ell_{max}}$ vanishes if $\ell_{max}\rightarrow \infty$. Therefore, the assumption $\delta_3=0$ implies some approximation. 

Another obvious relation comes from the observation that the sum of the four equations (\ref{Leq0}) is equal to zero. This in turn implies $\sum_0^3 (f_{\ell} + g_{\ell}+ f'_{\ell} + g'_{\ell})=0$. In \ref{Lan-Cen} we also show that this relation is valid only in the case of an infinite number of terms. However, this new relation is not relevant for our analysis because it involves the Landau parameter $f_0$. One should remind that no density-dependent terms have been included up to now, which affects in particular $f_0$ through the rearrangement contribution. In principle, applying the method of \cite{car08,rai11,dav13} to three-, four-, ... body interactions, a density-dependent pseudopotential would be obtained, which implies a density dependence of all Landau parameters, {\it i.e.} all channels $(\alpha)$ and all multipoles $\ell$. We shall follow however a simpler route and include the standard Skyrme effective density-dependent term $\frac{1}{6} t_3 (1+x_3 P_{\sigma}) \rho^{\gamma}$. It seems reasonable to fix beforehand the value of $\gamma$,  as it is usual in the fitting procedures, so that two more unknowns, $t_3$ and $x_3$, are introduced. This density-dependent term only affects the monopolar Landau parameters, to which we must add the contributions: 
%%%
$$
\left. \begin{array}{rl}
f_0(t_3) =& \frac{1}{16} (\gamma+1) (\gamma+2) t_3 \rho^{\gamma}\, , \\
g_0(t_3) =& - \frac{1}{24} (1-2x_3) t_3 \rho^{\gamma} \, ,\\
f'_0(t_3) =& - \frac{1}{24} (1+2x_3) t_3 \rho^{\gamma}\, , \\
g'_0(t_3) =& - \frac{1}{24}  t_3 \rho^{\gamma}  \,.
 \end{array} \right. 
$$
%%%
One can see that $g_0(t_3) + f'_0(t_3) = 2 g'_0(t_3)$, so it does not modify the condition (\ref{eq-SNM}). We are thus left with the system:
%%%
\begin{equation} \left.
\begin{array}{rl}
3 t_0 + \frac{1}{4} (\gamma+1) (\gamma+2) t_3 \rho^{\gamma} =& 4 \sum_0^3 f_{\ell}\, , \\
- (1-2x_0) t_0 - \frac{1}{6} (1-2x_3) t_3 \rho^{\gamma} =& 4 \sum_0^3 g_{\ell} \, ,\\
- (1+2x_0) t_0 - \frac{1}{6} (1+2x_3) t_3 \rho^{\gamma} =& 4 \sum_0^3 f'_{\ell}\, , \\
-  t_0 - \frac{1}{6}  t_3 \rho^{\gamma} =& 4 \sum_0^3 g'_{\ell} \, ,
 \end{array} \right\}
 \label{params-SNM}
 \end{equation}
 %%%
for four unknowns, but it actually reduces to three linearly independent equations due to (\ref{eq-SNM}).   

Up to now we have been considering only symmetric nuclear matter. 
Including pure neutron matter in our analysis provides {\it a priori} two new relations for the pseudopotential parameters. However, the one involving the spin $g_{\ell}^{(n)}$ parameters is in fact the sum of those involving $g_{\ell}$ and $g'_{\ell}$ in (\ref{params-SNM}). Only the relation involving $f_{\ell}^{(n)}$ matters, because of the rearrangement contributions. It reads:
%%%
\begin{equation} 
 (1-x_0) t_0 + \frac{1}{12} (\gamma+1) (\gamma+2)  (1-x_3) t_3 \rho^{\gamma} = 2 \sum_0^3 f^{(n)}_{\ell} \, .
 \label{params-PNM}
 \end{equation}
 
In principle Eqs. (\ref{params-SNM}) and (\ref{params-PNM}) allow us to determine $t_0$, $t_3$, $x_0$ and $x_3$ from the Landau parameters, after eliminating one of the three last equations of the set (\ref{params-SNM}).  
As far as equality (\ref{eq-SNM}) is satisfied the solution is unique, no matter which one of them is eliminated. 
For instance, dropping the equation involving the parameters $g'_{\ell}$ one gets the solutions:
%%% 
\begin{eqnarray} 
\fl t_0 &=& \frac{2}{3 \gamma (\gamma+3)} \sum_0^3 \left\{ -4  f_{\ell} - 3 (\gamma+1)(\gamma+2) ( f'_{\ell}+g_{\ell}) \right\} \, ,\\
\fl t_0 x_0 &=& \frac{2}{3 \gamma (\gamma+3)} \sum_0^3\left\{ -4  f_{\ell} -\frac{ 3}{2} (\gamma+1)(\gamma+2)  \left( f'_{\ell} - g_{\ell} \right) + 6  f^{(n)}_{\ell} \right\}\, ,\\
\fl \frac{1}{6} t_3 \rho^{\gamma}&=& \frac{2}{3 \gamma (\gamma+3)}\sum_0^3 \left\{ 4  f_{\ell} + 6(  f'_{\ell}+g_{\ell})  
\right\}\, , \\
\fl \frac{1}{6} t_3  x_3 \rho^{\gamma}&=&\frac{2}{3 \gamma (\gamma+3)} \sum_0^3\left\{ 4  f_{\ell} 
+3  \left( f'_{\ell} - g_{\ell} \right) -6  f^{(n)}_{\ell} \right\}\, .
\label{choixA}
 \end{eqnarray}
 %%%
The actual value of $\delta_3$ gives an estimate of the accuracy of the parameters so calculated. 

\subsection{Tensor parameters}
Proceeding along the same lines as in the previous subsection, the tensor Landau parameters can be written as:
%%%
\begin{equation} \left.
\begin{array}{rl}
h_0^{(\alpha)}  =&   \frac{1}{8} k_F^2 L_2[h^{(\alpha)} ] + \frac{1}{2} k_F^4 L_4[h^{(\alpha)} ] + \frac{1}{3} k_F^6 L_6[h^{(\alpha)} ] \, , \\ 
h_1^{(\alpha)}  =&  - \frac{1}{2} k_F^4 L_4[h^{(\alpha)} ]  - \frac{1}{2} k_F^6 L_6[h^{(\alpha)} ]\, ,  \\
h_2^{(\alpha)}  =&   \frac{1}{6} k_F^6 L_6[h^{(\alpha)} ]\, ,
 \end{array} \right\}
 \label{h0-2}
 \end{equation}
 %%%
where the combinations of pseudopotential parameters are:
%%%
\begin{equation} \left.
\begin{array}{rl}
L_n[h] =& t_e^{(n)} + 3 t_o^{(n)}\, ,  \\ 
L_n[h'] =&  - t_e^{(n)}+t_o^{(n)}\,.
 \end{array} \right\}
  \end{equation}
 %%%
Analogously to the central parameters for $n \ge 2$ the systems are invertible and we get~:
%%%
\begin{eqnarray}
\fl  \frac{2}{3} k_F^6 t_e^{(6)} &=& h_2 - 3 h'_2 \, ,\\
\fl  \frac{2}{3} k_F^6 t_o^{(6)} &=&  h_2 + h'_2   \, ,
\label{tenseur1}
\end{eqnarray}
%%%
\begin{eqnarray}
\fl - 2 k_F^4 t_e^{(4)} &=& h_1 + 3 h_2 - 3 h'_1 - 9 h'_2 \, ,\\
\fl - 2 k_F^4 t_o^{(4)} &=&  h_1 + 3 h_2 + h'_1 +  3 h'_2\, ,
\end{eqnarray}
%%%
\begin{eqnarray}
\fl k_F^2 t_e & = &  2 \sum_{\ell=0}^2 (h_{\ell} - 3 h'_{\ell}), \\
\fl k_F^2 t_o & = & 2 \sum_{\ell=0}^2 (h_{\ell} +h'_{\ell})\, .
\label{tenseur2}
\end{eqnarray}
%%%
This completes the determination of the tensor parameters. The use of PNM equations does not provide any new relation. 

\section{Two sets of N$^3$LO parameters}\label{sect-results}

To illustrate our exploratory analysis, we have calculated the Landau parameters using the phenomenological finite-range interactions D1MT \cite{gor09,ang11} and M3Y-P2 \cite{nak03}. The first one belongs to the Gogny family, supplemented with a tensor term related to the Argonne AV8' interaction, with a regularization term whose parameters have been fitted to some selected nuclear excited levels~\cite{ang11}. The second interaction is a superposition of Yukawa radial functions with different ranges, and includes tensor terms. All the parameters of this interaction have been obtained from a complete fitting procedure. Both interactions include a standard Skyrme density-dependent term. 

\begin{figure}[!h]
\begin{center}
\includegraphics[width=0.38\textwidth,angle=-90]{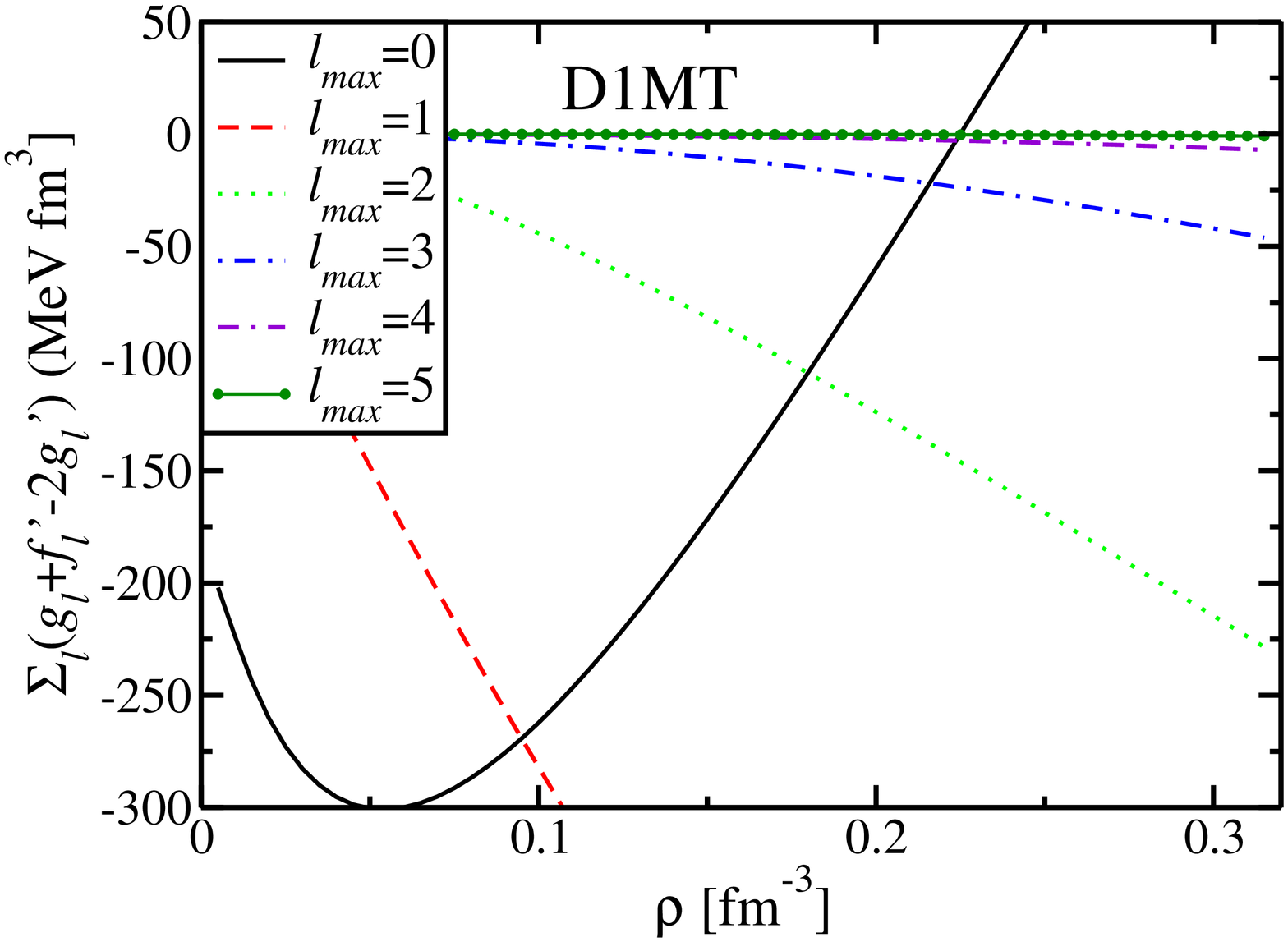}
\includegraphics[width=0.38\textwidth,angle=-90]{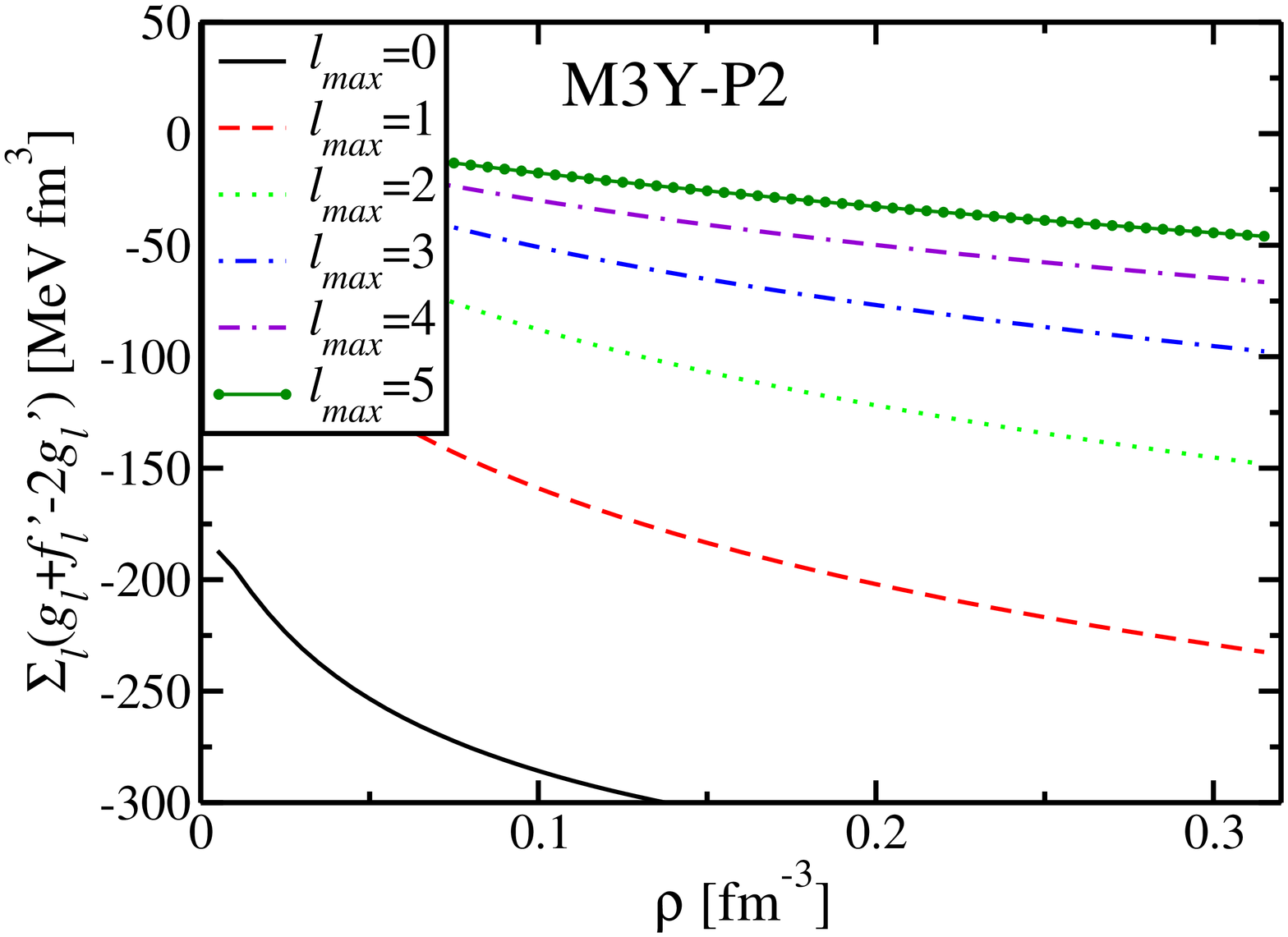}
\end{center}
\caption{(Color online). The quantity $\delta_{\ell_{max}}$ as given by Eq. (\ref{eq-SNM}) is plotted as a function of the density for increasing values of $\ell_{max}$ and for interactions D1MT \cite{gor09,ang11} and M3Y-P2 \cite{nak03}.}
\label{fig:diff}
\end{figure}

The should first check the validity of Eq. (\ref{eq-SNM}), since $\delta_3$ provides a hint about the unicity of the solutions and a rough error bar estimate of them as well. In Figure \ref{fig:diff} the quantity $\delta_{\ell_{max}}$ is plotted as a function of the density for values of $\ell_{max}$ from 0 to 5. As  $\ell_{max}$ increases, this quantity goes to zero for any value of the density. However a large density dependence is apparent for small values of  ${\ell}_{max}$. For instance, taking as a reference the density value 0.16 fm$^{-3}$ the quantity $\delta_1$ amounts to be about -420 and -190 MeV fm$^3$, respectively for D1M and M3Y-P2. The corresponding values for $\delta_3$ are -12 and -60 MeV fm$^3$. It is thus hopeless to fit the N$^1$LO parameters, but there are comfortably narrow bounds to start with the determination of the N$^3$LO ones. With this proviso in mind, we have determined up to the N$^3$LO parameters from these interactions at saturation density. Their values are given in Tables \ref{fits-central} and \ref{fits-tensor}, respectively for the central and tensor parameters. Obviously, since the aim is to obtain only an estimation, all the numerical values have been rounded off.

\begin{table}[h!]
\begin{center}
\begin{tabular}{l|ccc|ccc|}
\cline{2-7}& \multicolumn{3}{c|}{D1MT} & \multicolumn{3}{c|}{M3Y-P2}  \\
\cline{2-7}& N$^1$LO & N$^2$LO & N$^3$LO & N$^1$LO & N$^2$LO & N$^3$LO  \\
\hline
$t_0$ [MeV\,fm$^3$]        &  -283.5 & -1411  & -1670  & -1352 & -1640  & -1748     \\
$x_0$                                &  3.73 &  0.59 & 0.45      & 0.512 & 0.413  & 0.387     \\
$t_3$ [MeV\,fm$^{3+3\gamma}$]   &  56 & 7406  &  9115     &  6441  & 7520  &  7827   \\
$x_3$                                &  147.8 & 1.74  & 0.49            &  0.839 & 0.667  & 0.635          \\
\hline
$t_1$ [MeV\,fm$^5$]        &  305 & 537  & 617       & 336 & 716  & 1011      \\
$x_1$                                &  0.558 & 0.504  &  0.490  & 0.282  &  0.223 &  0.186    \\
$t_2$ [MeV\,fm$^5$]        &  -282 & 608  &   1031     & 124  & 378  &   689     \\
$x_2$                                &  -1.04 & -1.29  & -1.26    & -1.07  & -1.03  & -0.974    \\
\hline
$t_1^{(4)}$ [MeV\,fm$^7$]    &  - & -64.0 &  -119      &  - & -107 &  -314    \\
$x_1^{(4)}$                           &  -  & 0.43 &  0.42      &  -  &0.172 &  0.121      \\
$t_2^{(4)}$ [MeV\,fm$^7$]    & -   & -245  &   -537    & -   &  -71.3 &   -290     \\  
$x_2^{(4)}$                            &  -  & -1.22&  -1.21      &  -  & -1.03 &  -0.929    \\
\hline
$t_1^{(6)}$ [MeV\,fm$^9$]      &  -  &   -  & 2.52         &  -  &   -  &  9.75       \\
$x_1^{(6)}$                              &  -  &  -   &   0.39      &  -  &  -   &   0.0948   \\
$t_2^{(6)}$ [MeV\,fm$^9$]        &  -  &  -   &  13.4           &  -  &  -   &  10.3      \\
$x_2^{(6)}$                            &  -  &  -   &  -1.21      &  -  &  -   &  -0.901     \\
\hline
\end{tabular}
\caption{Central N$^3$LO parameters derived from Eqs. (\ref{choixA}), at density $\rho=0.16$fm$^{-3}$ and using the value $\gamma=1/3$, for interactions D1MT and M3Y-P2.}
\label{fits-central}
\end{center}
\end{table}

From  Table~\ref{fits-central} one can see that all the standard Skyrme parameters are significantly modified when going from N$^1$LO to N$^3$LO: the contribution of 4th- and 6th-order modifies them in a non negligible way, particularly in the D1MT case.  It is interesting to compare the coefficients $t_{1,2}^{(n)}$ in the N$^3$LO column.
Rigorously, one should multiply $t_i^{(4)}$ ($t_i^{(6)}$)  by $k_F^2$ ($k_F^4$) to have the same units as $t_{1,2}$. Keeping in mind that these values have been obtained for 
$k_F\simeq 1.3$ fm$^{-1}$ one can observe that 4th (6th) order parameters are one order of magnitude smaller than 2th- (4th-) order parameters, so that this expansion converges rapidly~\cite{car10}.

The tensor parameters have been estimated from Eqs. (\ref{tenseur1}-\ref{tenseur2}). The results are presented in Table \ref{fits-tensor}. The same type of conclusions as for the central parameters can be drawn for the tensor ones. 
One should keep in mind that, contrarily to M3T-P2, the D1MT tensor terms have been fixed to some levels and afterward added to the previously fitted central ones. In this sense, the numbers are probably less reliable for D1MT than for M3Y-P2.

\begin{table}[h!]
\begin{center}
\begin{tabular}{c|ccc|ccc|}
\cline{2-7}& \multicolumn{3}{c|}{D1MT} & \multicolumn{3}{c|}{M3Y-P2} \\
\cline{2-7}& Skyrme & N$^2$LO & N$^3$LO & Skyrme & N$^2$LO & N$^3$LO  \\
\hline
\hline
$t_e$  [MeV\,fm$^5$]& 140  & 412 &  670 & 23   & 54  & 76 \\
$t_o$  [MeV\,fm$^5$]&  47 & 137 &  223 & 7  & 19 & 27  \\
\hline
$t_e^{(4)}$  [MeV\,fm$^7$]& - & -38 &  -145 & -   &  -4 & -14 \\
$t_o^{(4)}$  [MeV\,fm$^7$]& - &  -13 &  -48 & -  &  -2 & -5 \\
\hline
$t_e^{(6)}$ [MeV\,fm$^9$] & - & - &  60 & -   & -  & 5 \\
$t_o^{(6)}$  [MeV\,fm$^9$]& - & - &  20 & -  & - & 2 \\
\hline
\end{tabular}
\caption{Tensor N$^3$LO parameters derived from interactions D1MT and M3Y-P2.}
\label{fits-tensor}
\end{center}
\end{table}

\begin{figure}[!h]
\begin{center}
\includegraphics[width=0.6\textwidth,angle=-90]{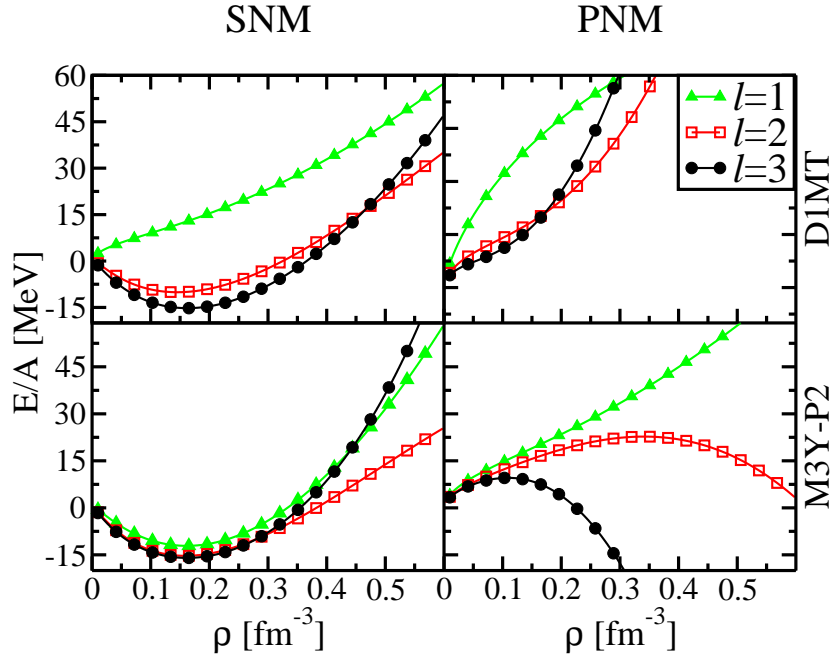}
\end{center}
\caption{(Color online) Energies per particle in symmetric nuclear matter (left panels) and pure neutron matter (right panels) as obtained in N$^{\ell}$LO for interactions D1MT (upper panels) and M3Y-P2 (lower panels).}
\label{fig:eos}
\end{figure}

The density value $\rho=0.16$ fm$^{-3}$ has been fixed to determine the parameters. To guess whether they give reasonable physical results or not, we have computed the energy per particle for symmetric nuclear matter (SNM) and pure neutron matter (PNM) as a function of density:
\begin{eqnarray}
\fl E/A \bigg|_{SNM}  & = & \frac{3}{5} \frac{\hbar^2}{2 m} k_F^2 + \frac{3}{8} t_0 \rho + \frac{1}{16} t_3 \rho^{1+\gamma}
+\frac{3}{80}  \left\{ 3 t_1 + (5+4x_2) t_2 \right\} \rho k_F^2  \nonumber \\
& + &  \frac{9}{280} \left\{ 3 t_1^{(4)} + (5+4x_2^{(4)}) t_2^{(4)} \right\} \rho k_F^4 
+  \frac{2}{15} \left\{ 3 t_1^{(6)} + (5+4x_2^{(6)}) t_2^{(6)} \right\} \rho k_F^6 \\
\fl E/N \bigg|_{PNM}  & = & \frac{3}{5} \frac{\hbar^2}{2 m} k_F^2 + \frac{1}{4} t_0 (1-x_0) \rho + \frac{1}{24} t_3 (1-x_3) \rho^{1+\gamma}  \nonumber \\
&+& \frac{3}{40}  \left\{ t_1 (1-x_1) + 3 t_2 (1+x_2) t_2 \right\} \rho k_F^2  \nonumber \\
& + & \frac{9}{140} \left\{  t_1^{(4)} (1-x_1^{(4)})+3 t_2^{(4)} (1+x_2^{(4)})  \right\} \rho k_F^4  \nonumber \\
 &+&   \frac{1}{15} \left\{ 3 t_1^{(6)}(1-x_1^{(6)})+ 3 t_2^{(6)} (1+x_2^{(6)}) \right\} \rho k_F^6 
\end{eqnarray}
These quantities are plotted in Figure \ref{fig:eos} for the three successive approaches 
N$^{\ell}$LO, $\ell=1, 2, 3$. 

There are significant differences between the results obtained from the two employed interactions. Concerning SNM (left panels), the N$^1$LO parameters derived from D1MT do not bind SNM, and the two next orders slowly converge to a reasonable equation of state. In contrast, M3Y-P2 results provide a nice convergence.  
Looking now PNM (right panels) one can see that M3Y-P2 tends to a collapse as density is increased, but D1MT gives 
a seemingly reasonable equation of state. Obviously these results show how the parameters are relied to the  interaction used as an input. It should be noticed that the original interaction M3Y-P2 does not collapse~: this is due of course to the fact we truncate at 3rd order. This indicates the delicate balance for parameters and pointed out that a real fitting procedure has to be done beyond our starting point.

To be more quantitative, in Table \ref{results} we present the energy per particle $\mathcal{E}_{0}$ and the incompressibility $\mathcal{K}_{0}$,
calculated at the saturation density $\rho_0$ in SNM. We observe that the complete set of interaction parameters obtained from M3Y-P2 Landau parameters give results that are pretty close to the original one. This is in contrast with the results obtained with the partial set (when one truncates at N$^1$LO or N$^2$LO). 
\begin{table}[h!]
\begin{center}
\begin{tabular}{c|ccc|ccc|}
\cline{2-7}& \multicolumn{3}{c|}{D1MT} & \multicolumn{3}{c|}{M3Y-P2}  \\
\cline{2-7}        & N$^1$LO & N$^2$LO & N$^3$LO & N$^1$LO & N$^2$LO & N$^3$LO\\
\hline
$\rho_{0}$ [fm$^{-3}$]                  & - & 0.145 & 0.163     &0.161 & 0.1659 & 0.162   \\
$\mathcal{E}_{0}$ \,\,\,[MeV]           &  unbound   & -10.1 & -15.2     &-12.1  & -15.3  & -16.0      \\
$\mathcal{K}_{0}$  \,\,[MeV]           &  - &  154 & 216   & 213   &   207   &  217      \\
\hline
\end{tabular}
\caption{Numerical results for saturation properties in infinite nuclear matter obtained with the two sets of N$^3$LO parameters determined from D1MT and M3Y-P2.}
\label{results}
\end{center}
\end{table}

\section{Summary and conclusions}\label{sect-conclusions}

We have shown a possible way to estimate the N$^3$LO pseudopotentials by using Landau parameters derived from finite-range interactions. The method permits to fix the central $t_1^{(n)}, x_1^{(n)}, t_2^{(n)}, x_2^{(n)}$ and tensor $t_e^{(n)}, t_o^{(n)}$ parameters at all orders of approximation. However, the determination of the zeroth-order parameters $t_0, x_0$ is not unique, because it requires the Landau parameters to fulfill some relations which depend on the considered order. We have identified the origin of the problem to be the absence of density dependent terms in the  pseudopotential. In general, it would include three-, fourth-, \dots body terms, which will generate a density dependence. We have partly remedy to this problem in a simple way, by including a standard Skyrme density-dependent term. This introduces two new parameters, $t_3, x_3$, assuming a fixed value for the power $\gamma$ of the density. The system of equations to get the four zeroth-order parameters is still overdetermined, as far as the sum $\delta_3$ is different from zero. The quantity $\delta_3$ (see Eq. \ref{eq-SNM}) involves Landau spin, isospin and spin-isospin parameters, and its departure from zero provides thus a rough estimate of the uncertainties in the zeroth-order parameters.  
 
This procedure requires the knowledge of Landau parameters calculated in a consistent way, both for symmetric nuclear matter and pure neutron matter, in a wide range of densities around the saturation density value of symmetric nuclear matter. To the best of our knowledge, no calculations based on realistic interactions with all these requirements exist in the literature.  We have thus made estimates of the N$^3$LO pseudopotential parameters based on the phenomenological interactions D1MT and M3Y-P2, from which we have calculated the required Landau parameters.  
We have checked that the values of $\delta_3$ lead to reasonable errors. We have tested the obtained central parameter sets, showing that they reasonably reproduce the equation of state of symmetric nuclear matter. As should be expected, the results rely on the  interaction used as an input. The use of Landau parameters consistently derived from realistic interactions would be of a great help to fix the starting point for obtaining the N$^3$LO pseudopotential.  
To this respect, we must stress that the present method is not a new procedure to fit pseudopotential parameters. 
In fact, surface properties \cite{nik11} are not considered in this scheme. 
The interest of the method is that it can provide a good starting point for the usual fitting procedure, which must include a fit to finite nuclei properties. Work is in progress in that direction \cite{les_finlandais}.

\section*{Acknowledgments}
This work was supported Mineco (Spain), grant FIS2011-28617-C02-02. The authors thank J. Meyer for stimulating discussions.

\appendix

\section{Some relations between Landau parameters deduced from a general central interaction}
\label{Lan-Cen}
Consider a general two-body central interaction which we write as
\begin{eqnarray}
V(r_{12}) = W(r_{12}) + B(r_{12}) P_{\sigma} - H(r_{12}) P_{\tau} - M(r_{12})
P_{\sigma}  P_{\tau}  \, ,
\label{Vpp}
\end{eqnarray}
where $P_{\sigma} , P_{\tau} $ are the usual spin and isospin exchange operators. 
As it does not contains any density-dependence, the associated antisymmetrized particle-hole (ph) interaction is obtained by multiplying (\ref{Vpp}) with $1- P_x P_{\sigma}  P_{\tau} $, where $P_x$ is the space-exchange operator, and calculating its matrix element between ph states. Due to momentum conservation, there are at most three momenta, which we choose to the the initial (final) momentum ${\bf k}_1$ (${\bf k}_2$) of the holes and the momentum transfer ${\bf q}$. As a result, in each spin-isospin channel $(\alpha)$, the ph interaction can be written as~:
\begin{eqnarray*}
V_{ph}^{(\alpha)}(q,{\bf k}_1,{\bf k}_2) = D^{(\alpha)}(q) - E^{(\alpha)}({\bf k}_1-{\bf k}_2) \, .
\end{eqnarray*}
Denoting a Fourier transform with a tilde, and omitting the arguments, the direct and exchange terms are given by
\begin{eqnarray*}
D^{(0,0)} =  \tilde{W} + \frac{1}{2} \tilde{B} - \frac{1}{2} \tilde{H} - \frac{1}{4} \tilde{M}  & \quad , \quad & 
E^{(0,0)} =  \frac{1}{4} \tilde{W} + \frac{1}{2} \tilde{B} - \frac{1}{2} \tilde{H} -  \tilde{M}    \\
D^{(1,0)} =  \frac{1}{2} \tilde{B}  - \frac{1}{4} \tilde{M}   & \quad , \quad &
E^{(1,0)} =   \frac{1}{4} \tilde{W}  - \frac{1}{2} \tilde{H}     \\
D^{(0,1)} =  - \frac{1}{2} \tilde{H} - \frac{1}{4} \tilde{M}  & \quad , \quad &
E^{(0,1)} = \frac{1}{4} \tilde{W} + \frac{1}{2} \tilde{B}     \\
D^{(1,1)} =  - \frac{1}{4}  \tilde{M}  &\quad , \quad & 
E^{(1,1)} =  \frac{1}{4}  \tilde{W}.  
\end{eqnarray*}

In the Landau approximation particles and holes are assumed to be on the Fermi surface, that is $q=0$, $k_{1,2}=k_F$, so that the interaction only depends on $\cos \theta = (\hat{k}_1 \cdot \hat{k}_2)$. 
The argument of the direct term is thus zero, and that of the exchange term is $\sqrt{2 k_F^2 (1- \cos \theta)}$. 
The Landau parameters are the coefficients of an expansion of the particle-hole interaction in Legendre polynomials
For each $(\alpha)$ channel, the Landau parameters are defined as
\begin{eqnarray*}
V_{ph}^{(\alpha)} = \sum_{\ell} f_{\ell}^{(\alpha)} \, P_{\ell}( \cos \theta).
\end{eqnarray*}
We are looking for sums of the type $\sum_{\ell} f_{\ell}^{(\alpha)}$.  Since
$P_{\ell}( 1) = 1$, such sums are obtained taking the value $\cos \theta = 1$ in the ph interaction.
We thus have~:
\begin{eqnarray*}
\sum_{\ell} f_{\ell}^{(\alpha)} =  V^{(\alpha)}_{ph}(\cos \theta = 1) = D^{(\alpha)}(0) - E^{(\alpha)}(0)\,.
\end{eqnarray*}
One immediately checks that the equations
\begin{eqnarray*}
 && \sum_{\ell} \left( g_{\ell} + f'_{\ell} - 2 g'_{\ell} \right) = 0 \, ,\\ 
 && \sum_{\ell} \left( f_{\ell} + g_{\ell} + f'_{\ell} + g'_{\ell} \right) = 0 \, ,
\end{eqnarray*}
are actually identities.


\section*{References}
\begin{thebibliography}{99}
\bibitem{ben03} M. Bender, P-H. Heenen and P-G. Reinhard, Rev. Mod. Phys. {\bf 75}, 121 (2003).
\bibitem{kor10}M. Kortelainen, T. Lesinski, J. Mor\'e, W. Nazarwicz, J. Sarich, N. Schunck, M.V. Stoitsov and S. Wild, Phys. Rev. C {\bf 82}, 024313, (2010).
\bibitem{kor13}M. Kortelainen, J. McDonnell, W. Nazarewicz, E. Olsen, P.-G. Reinhard, J. Sarich, N. Schunck, S.M. Wild, D. Davesne, J. Erler, A. Pastore,  arXiv:1312.1746
\bibitem{sad13}J. Sadoudi, T. Duguet, J. Meyer, and M. Bender, Phys. Rev. C {\bf 88}, 064326 (2013)
\bibitem{gor13}S. Goriely, N. Chamel, and J. M. Pearson, Phys. Rev. C {\bf 88}, 061302 (2013)
\bibitem{was12}K. Washiyama, K. Bennaceur, B. Avez, M. Bender, P.-H. Heenen, and V. Hellemans, Phys. Rev. C {\bf 86}, 054309 (2012)
\bibitem{car08} B.G. Carlsson, J. Dobaczewski and M. Kortelainen,  Phys. Rev. C {\bf 78}, 044326 (2008).
\bibitem{rai11} F. Raimondi, B. G. Carlsson and J. Dobaczewski, Phys. Rev. C {\bf 83}, 054311 (2011).
\bibitem{car10} B. G. Carlsson and J. Dobaczewski, Phys. Rev. Lett. {\bf 105}, 122501, (2010).
\bibitem{sto10} M. Stoitsov, M. Kortelainen, S. K. Bogner, T. Duguet, R. J. Furnstahl, B. Gebremariam, and N. Schunck, Phys. Rev. C {\bf 82}, 054307 (2010)
\bibitem{dav13} D Davesne, A Pastore and J Navarro, J. Phys. G: Nucl. Part. Phys. {\bf 40} 095104 (2013).
\bibitem{rai11b} F. Raimondi, B. G. Carlsson, J. Dobaczewski, and J. Toivanen,  Phys. Rev. C {\bf 84}, 064303 (2011).
\bibitem{doba95} J. Dobaczewski, J. Dudek, Phys. Rev. C{\bf 52} (1995) 1827; 
Phys. Rev. C {\bf 55}, 3177 (1997).
\bibitem{bla86} J.P. Blaizot and G. Ripka, {\em Quantum Theory of Finite Systems}, MIT Press, Cambridge (1980). 
\bibitem{vau72}D. Vautherin and D. M. Brink, Phys. Rev. C {\bf 5}, 626 (1972)
\bibitem{cha98} E. Chabanat, P. Bonche, P. Haensel, J. Meyer, and R. Schaeffer, Nucl. Phys. {\bf A 635}, 231 (1998)
\bibitem{cha98b} E. Chabanat, P. Bonche, P. Haensel, J. Meyer, and R. Schaeffer,  Nucl. Phys.{\bf A 643}, 441 (1998). 
\bibitem{gor09} S. Goriely, S. Hilaire, M. Girod, and S. P\'eru, Phys. Rev. Lett. {\bf 102}, 242501 (2009). 
\bibitem{ang11} M. Anguiano, G. C\'o, V. De Donno, and A.M. Lallena, Phys. Rev. C {\bf 86}, 054302 (2012).
\bibitem{nak03}H. Nakada,  Phys. Rev. C{\bf 68}, 014316 (2003).
\bibitem{dob10} J. Dobaczewski, B.G. Carlsson and M. Kortelainen,  J. Phys. G: Nucl. Part. Phys. {\bf 37}, 075106 (2010).
\bibitem{nik11}N. Nikolov, N. Schunck, W. Nazarewicz, M. Bender, and J. Pei, Phys. Rev. C {\bf 83}, 034305 (2011).
\bibitem{les_finlandais} J. Dobaczewksi and G. Carlsson, private communication.
\end{thebibliography}
\end{document}